\documentclass[conference]{IEEEtran}
\ifCLASSINFOpdf
\else
\fi
\usepackage{booktabs}
\usepackage{dcolumn}
\usepackage{units}
\usepackage{array}
\usepackage{tabu}
\usepackage{booktabs}

\usepackage[table]{xcolor}
\usepackage{amsmath}
\usepackage{multirow, balance}
\usepackage{amsmath}
\usepackage{algorithm}
\usepackage[noend]{algpseudocode}

\usepackage{amsmath}
\usepackage{algpseudocode}
\usepackage{array}
\usepackage{algpseudocode}
\usepackage[pdftex]{graphicx}
  \graphicspath{{./fig/}}
  \usepackage{amssymb}
\usepackage{pifont}
\usepackage{tabularx}
\usepackage{tabulary} 
\usepackage{rotating}
\usepackage{varwidth}
\usepackage{amsmath}
\usepackage{algorithm}
\usepackage[noend]{algpseudocode}
\usepackage{listings}

\makeatletter
\def\BState{\State\hskip-\ALG@thistlm}
\makeatother

\begin{document}

\newcolumntype{f}{>{$}l<{$}}
\newcolumntype{n}{l}
\newcolumntype{N}{>{\scriptsize}l}
\newcolumntype{v}[1]{>{\raggedright\hspace{0pt}}p{#1}}
\newcolumntype{V}[1]{>{\scriptsize\raggedright\hspace{0pt}}p{#1}}
%
\newcolumntype{B}[1]{>{\boldmath\DC@{.}{,}{#1}}l<{\DC@end}}
\newcolumntype{d}[1]{>{\DC@{.}{,}{#1}}l<{\DC@end}}
\newcolumntype{i}[1]{>{\DC@{.}{,}{#1}\mathnormal\bgroup}l<{\egroup\DC@end}}
\newcolumntype{s}[1]{>{\DC@{.}{,}{#1}\mathsf\bgroup}l<{\egroup\DC@end}}
\newcolumntype{L}[1]{>{\raggedright\arraybackslash}p{#1}} 
\newcommand\tabrotate[1]{\begin{turn}{90}\rlap{#1}\end{turn}}
\newcommand\tabvarwidth[2][3cm]{\begin{varwidth}[b]{#1}\centering #2\end{varwidth}}

\newcolumntype{L}[1]{>{\raggedright\arraybackslash}p{#1}} 
\newcolumntype{C}[1]{>{\centering\arraybackslash}m{#1}}
%
\title{Software-Based Fault Recovery via  \\ Adaptive Diversity for COTS Multi-Core Processors}

\author{\IEEEauthorblockN{Andrea H{\"o}ller, Tobias Rauter, Johannes Iber, Georg Macher and Christian Kreiner}
\IEEEauthorblockA{Institute for Technical Informatics, Graz University of Technology, Austria\\
\{andrea.hoeller, tobias.rauter, johannes.iber, georg.macher, christian.kreiner\}@tugraz.at
}
}


\maketitle

\begin{abstract}
The ever growing demands of embedded systems to satisfy high computing performance and cost efficiency lead to the trend of using commercial off-the-shelf hardware. However, due to their highly integrated design they are becoming increasingly susceptible to hardware errors (e.g. caused by radiation-induced soft-errors or wear-out effects). Since such faults cannot be fully prevented, systems have to cope with their effects. At the same time there is the trend of multi-core processors in embedded systems. Approaches to achieve fault tolerance by using the multiple cores to establish redundancy have been presented in literature. However, typically only homogeneous redundancy techniques are considered to tolerate soft errors.  However, there is a lack of appropriate reaction mechanisms for restoring the system in case of permanent hardware faults. 

Here, we propose the basic idea of enhancing multi-core redundancy techniques with a cost-efficient automated introduction of diversity in the executed software replicas. Recently, these automated software diversity techniques have attracted attention in the security domain. We propose to use these techniques to recover from permanent hardware faults. This is achieved by adapting the software execution in such a way that permanent faults are mitigated. 
\end{abstract}


%
\IEEEpeerreviewmaketitle

\section{Introduction}
Cyber-physical systems (CPS) play a central role in many critical application domains such as industrial control, automotive, and healthcare. Consequently, the reliability and availability of these systems has become a significant concern. At the same time cyber-physical systems are facing increasing demands on high computing performance (e.g., for autonomous driving). This leads to a move to commercial off-the-shelf (COTS) hardware allowing designers to use state-of-the-art components that guarantee a high performance \cite{Alhakeem2015}. However, the vulnerability to hardware faults increases due to the continuing structure downscaling of the semiconductor industry. Thus, soft errors due to radiation and permanent errors due to manufacturing, process variations, ageing, etc. are expected to occur more often in future \cite{Saggese2005}.

Established solutions to provide the required reliability are based on hardened components. However, such hardware-based approaches are very costly. Furthermore, commercially-available processors that are especially designed and manufactured for safety-critical applications typically only offer limited computing performance features. Therefore, software-based approaches for error detection and correction are becoming increasingly important. Furthermore, with the advent of mult-core technology, hardware redundancy techniques, which typically impose very high hardware overheads, are becoming ever more attractive \cite{Gizopoulos2011}. For example, a Triple Modular Redundancy (TMR) mechanism can be implemented by triplicating a critical task on three cores and comparing the results. 

Although much research in recent years has focused on enhancing the reliability of multi-core systems with redundancy, they typically only consider homogeneous redundancy (e.g. \cite{Alhakeem2015,Gizopoulos2011,Dobel2012}). This means that the same binary is replicated on the redundant cores. Considering a TMR system, this allows to detect transient and permanent faults in one core. Furthermore, permanent faults in one core can be masked, since the remaining two cores produce the correct result. Typical solutions for many core systems (e.g. highly parallel processors) involve deactivating the faulty core and shifting the calculation to a spare core \cite{Alhakeem2015}. Due to the limited resources of embedded systems (i.e. power, production costs) they are composed of multi-core processors with a low number of cores (i.e. less than eight cores). Consequently, few spare cores are available, which limits the means of these fail-over concepts. Thus, in order to increase the availability, it is desired to recover from an identified permanent malfunctioning of a core.

Here, we present a software-based approach of establishing heterogeneous redundancy on a multi-core platform. We propose to push multi-core redundancy techniques towards the next level by introducing diversity in the redundant executions in a cost-efficient way. The main idea is to exploit concepts of research on automated software diversity that has mainly been published in the security domain to increase the fault recovery capabilities of redundancy implemented on a multi-core system. We suggest a high-level concept of using capabilities of these techniques to dynamically change the diversity in execution to recover from detected faulty states. The main idea is to adapt the execution of the diverse program replicas in such a way that a permanent hardware fault is bypassed. 

\section{Background and Related Work}

\subsection{Self-Healing Systems}
In order to deal with unforeseen events the idea of software self-adaptability has received attention \cite{Brun2009}. For example, self-healing systems autonomously detect and recover from faulty states by changing their configuration. However, so far these techniques are mainly used in complex server systems. Methods for embedded systems to recover from an unhealthy state are still a research challenge. Although hardware faults can be bypassed with self-modifying hardware (e.g. \cite{Jafri2010}), this technique is not applicable for typical COTS hardware and only offers limited flexibility. Thus, there remains the need for sophisticated software-based methods to handle unforeseen scenarios caused by faults. 

\subsection{Fault Tolerance via Multi-Core Redundancy}
The current development trend for computing platforms has moved from increasing the frequency of single cores to increasing the parallelism with multiple cores on the same die \cite{Macher2015}. Although, multi-core technologies present new development challenges, they have strong potential to support a cost-efficient creation of fault tolerance. Several online error detection techniques for multi-core processor components have been proposed. According to \cite{Gizopoulos2011} these techniques can be classified in periodic built-in self-test, dynamic verification, anomaly detection and redundant execution approaches. Here, we only focus on the latter and use built-in self-tests as a supplementary technique. 

N-modular redundancy is a widely-used fault detection technique where multiple processing elements process the same data. A well-known example is Dual Modular Redundancy (DMR) using two elements. Consequently, an error is detected, if the results of the two elements are different. Another prominent technique is Triple Modular Redundancy (TMR) that can also identify one faulty element with majority voting. This leads to a higher availability, since the system can continue the execution by masking the faulty element \cite{Mushtaq2011}. Since the voter is a single point of failure, it has to be highly reliable. 
Furthermore, these techniques can be classified in spatial and temporal redundancy techniques. While spatial redundancy means that the calculation is performed on distinct hardware components, temporal redundancy indicates that the same calculation is performed multiple times one after the other. Typically, redundancy established on a multi-core realizes spatial redundancy exploiting the inherent replication of processor cores. The cost-efficient achievement of hardware redundancy on a multi-core system is supported by the fact that the full utilization of cores is usually not feasible and, therefore unused processor cores can be used to execute redundant threads. 

The two main hardware-based approaches of redundant execution that have been proposed in literature are the lockstep configuration and the redundant multithreading (RMT) with and without loose lockstepping \cite{Gizopoulos2011}. 
In typical lockstep approaches identical cores are tightly-coupled and cross-checks are performed at each cycle or instruction execution.


Compared to hardware-based redundancy techniques, software-based approaches provide a low-cost alternative, since they impose a smaller hardware overhead and offer a higher level of flexibility. 
Several approaches of how to realize software-based redundancy on a multi-core system have been proposed. The authors of \cite{Bolchini2012} propose the replication of a task on three cores and to execute the comparison on a core that offers a high reliability. In \cite{Dobel2012} Romain - an OS service providing an adaptively replicated execution - is presented. 

Related work considers to adapt the number of parallel redundant executions during runtime \cite{Dobel2012,Alhakeem2015}. We propose to further increase the adaptability by also adapting the way of how the software is executed. Most of the presented approaches use homogeneous redundancy to detect hardware faults during operation. However, if both variants are identical and include the same systematic fault, they fail in the same way and the fault is not detected. Thus, the reliability of a redundant system not only depends on the reliability of each version, but also on the dissimilarity between them. There have been proposals to use the diversity of a heterogenous multi-core architecture to increase the fault tolerance \cite{Ungerer2013}. However, these approaches only offer limited portability. Thus, we propose to realize the diversity not in hardware, but in software. In \cite{Gizopoulos2011} the idea of using diverse software-component replicas in a multi-core environment is presented. Here, we describe how to generate the diverse software replicas in an automated way.


\subsection{Automated Software Diversity}
Automated software diversity (ASD) techniques introduce diversity in the execution of the program. Recently, automated diversity gained attention in the security domain as a technique of diversifying each deployed program version \cite{Larsen2014}. This forces attackers to target each system individually. In contrast to N-version programming \cite{Pullum2001}, the diversity is introduced without manual interactions of humans in the loop. A common approach to realize such an automated software diversification is randomization to create "diversity in execution". This "diversity in execution" can denote diverse timings, diverse outputs, diverse memory usages etc. In contrast to static techniques that generate multiple diverse program versions, dynamic software diversity techniques use only one binary that is deployed and introduces the diversity during operation. More details about automated software diversity research is provided in \cite{Larsen2015,Baudry2014}.



\subsubsection{Dynamic Software Diversity}
\renewcommand{\arraystretch}{1.1}
\begin{table}[tb]
\centering
\small
\caption{Examples of adjustable parameters of dynamic software diversity methods}
\label{tab:params}
\begin{tabular}{L{5.5cm}L{2.4cm}}
\toprule
\textit{Randomization method}     & \textit{Parameter} \\
\midrule
Memory gaps between objects \cite{Bhatkar2005}       & Gap size         \\
Changing base address of program \cite{Bhatkar2005}  & Base address     \\ 
Changing base address of libraries and stack \cite{Chew2002}     & Base address \\
Permutation of the order of routine calls variables \cite{Bhatkar2005}    & Order of calls \\
Permutation of the order of variables \cite{Bhatkar2005} & Order of variables \\
Insertion of NOP instructions   \cite{Homescu2013}   & Number of NOPs   \\
Data re-expression / data diversity \mbox{(\textit{in’=f(in,k), out =} $f^{-1}$\textit{(out’,k)})} \cite{Amman1988h} & Parameter in re-expression \mbox{algorithm} (\textit{k})\\
\bottomrule
\end{tabular}
\vspace{0.5cm}
\centering
\caption{Examples of dynamic automated diversity techniques and the goals for which they have been proposed in literature.}

\small
\begin{tabular}{L{3.4cm}|C{0.2cm}C{0.2cm}C{0.2cm}|C{0.2cm}C{0.2cm}C{0.2cm}|C{0.2cm}C{0.2cm}}
\toprule
\multirow{5}{*}{\textit{Known Use}}       &
\multicolumn{3}{c|}{\tabvarwidth{\textit{Level of} \\ \textit{Diversific.}}} &
\multicolumn{3}{c|}{\tabvarwidth{\textit{Time of} \\ \textit{Diversific.}}} &
\multicolumn{2}{c}{\textit{Goal}} \\
& \multicolumn{3}{c|}{} & \multicolumn{3}{c|}{} &  \\ 
& \multicolumn{3}{c|}{} & \multicolumn{3}{c|}{} &  \\ 
& \multicolumn{3}{c|}{} & \multicolumn{3}{c|}{} &  \\ 
& \multicolumn{3}{c|}{} & \multicolumn{3}{c|}{} &  \\ 
 & 
\multirow{1}{*}{\tabrotate{\textit{Instruction}}} &
\multirow{1}{*}{\tabrotate{\textit{Function}}} &
\multirow{1}{*}{\tabrotate{\textit{Program}}}&
\multirow{1}{*}{\tabrotate{\textit{Installation}}} & 
\multirow{1}{*}{\tabrotate{\textit{Loading}}} & 
\multirow{1}{*}{\tabrotate{\textit{Execution}}} & 
\multirow{1}{*}{\tabrotate{\textit{Security}}} & 
\multirow{1}{*}{\tabrotate{\textit{Reliability}}}  \\ \midrule

Data randomization \cite{Amman1988h}        &    & x &   &    &   & x &  & x \\
Memory layout randomization (e.g. \cite{Bhatkar2005,Chew2002})  &    & x & x &  x &   &   & x &  \\
Program encoding randomization (e.g. \cite{Barrantes2005})          &  x &   &   &    &   & x & x &  \\
In-place diversification   (e.g. \cite{Pappas2012})     &    &   & x &    & x &   & x &  \\
Instruction location randomization  (e.g. \cite{Hiser2012,Scott2003}        &  x &   &   &    &   & x & x &  \\
Binary stirring \cite{Wartell2012}        &  x &   &   &    & x &   & x &  \\
\bottomrule
\end{tabular}
\label{tab:dynamicku}
\end{table}

Dynamic randomization methods create only one single version of an executable program that is able to perform its executions differently. These techniques either adapt the interfaces or the implementation \cite{Cowan2000}. Interface adaptions work on top of the code that is protected. They modify the layout or the interfaces, without changing the implementation of the core code that uses the interfaces. Implementation diversifications, do not alter the interface, but the implementations of portions of the code to make it resistant. Several approaches make programs self-randomizing by instrumenting them to mutate one or more implementation aspects as the program is loaded by the operating system or as it runs \cite{Larsen2014}. 

Dynamic automated software diversity is often realized by the operating system without the need to change the user program itself. However, appropriate mechanisms could also be built in the program as shown in \ref{tab:params}. Table \ref{tab:dynamicku} presents examples of dynamic ASD techniques presented in literature. Most of the techniques have been researched for security purposes. However, we assume that adaptations of these approaches may also be used to reach reliability goals. Dynamic reconfiguration could be applied in such a way that self-healing is established by bypassing detected faults. A well-established method that follows that principle is data re-expression. It transforms the original input to produce new inputs to redundant variants \cite{Amman1988h}. After execution the distortion introduced by the re-expression is removed before comparison. So a given initial data within the program failure region can be re-expressed to an input data that circumvents the faulty region \cite{Pullum2001}. 

\subsubsection{Static Software Diversity}
Static software diversity techniques automatically generate multiple diverse program variants based on the same source code before distributing the program. The two main approaches are to perform program transformation, where a diversified source code version is generated, or to introduce diversity during the compilation/linking stage. There are two main approaches. First, diversity could be introduced by performing code transformations meaning that based on a common source code basis diverse source code variants are generated. Second, the diversification could be done during compiling/linking by using one source code basis to generate diverse binaries.   

Additionally, it can be seen, that some techniques have been proposed in the security domain as well as in the reliability domain. This indicates the high potential cross-fertilization of these techniques. For example, a simple way of automatically introducing diversity in execution is to use different compilers and compiler options to generate multiple program versions. It has been shown that diverse compiling not only enhances the security of systems \cite{Wheeler2009} but it can also increase the hardware and software fault tolerance \cite{Gaiswinkler2009,Hoeller2015,Hoeller2015dasc}.

However, there are also many examples of static ASD techniques that have only been investigated to fulfill either security or reliability goals. Most techniques targeting the security are used to diversify a software program before deploying it on different targets \cite{Larsen2014}. However, we expect that the basic concepts of these techniques also have a high potential of increasing the overall resilience (e.g. hardware-fault tolerance) in a redundant configuration on a multi-core system. 

We propose to execute diverse replicas on a multi-core system to recover from detected faults. The above described techniques represent possible methods that can be applied to create these diverse replicas. These techniques introduce diversity in execution, which means that diverse software executions will show different timing characteristics and/or use different memory cells or processor instructions. The goal of diversification methods for fault recovery is that the faulty region is bypassed from the adapted execution.

\section{Fault Recovery Approach}
\subsection{System Model}
The target applications for our techniques are cyber-physical systems, where an embedded device is used to perform some control application. Such devices typically operate by performing an input acquisition to process data coming from sensors. Then, these input data is processed and the output is propagated to physical actuators. Here, we focus on how to increase the reliability of the processing task. 

The considered embedded device consists of a multi-core processor. We assume that software-based mechanisms to manage a redundant calculation of important tasks is already in place (e.g. as described in \cite{Alhakeem2015,Doebel2012,Ulbrich2012}). Thus, tasks with high reliability requirements are executed redundantly on multiple cores. After all calculations finished, a separate core executes an acceptance test comparing the results of the redundant calculations with a majority voter. Since this voting involves a single point of failure, it has to be protected. Techniques to achieve this have been proposed in previous work. For example, in \cite{Alhakeem2015} it is proposed to use software-based self-tests to ensure that the needed components to execute the voting procedure are function successfully at certain checkpoints. In the fault free case, the overhead of such tests is relatively low, since only those components that are used by the voting mechanism have to be checked.

We assume a \textit{N}-modular redundant system, meaning that \textit{N} identical modules are synchronized in order to perform voting of the outputs. The voting scheme is implemented as a basic \textit{M}-out-of-\textit{N} majority voting. Thus, $\lceil N/2 \rceil$ is the minimum number of modules free from error that are required to get a majority count. Thus, the technique allows at most \textit{N-M} cores to fail. Additionally, we assume that all modules fail independently. For the sake of simplicity, we use a TMR system with a \textit{2oo3} majority voting as an example in the remaining paper. Such a TMR system can tolerate faults in one core and the faulty core is identified.

Furthermore, we are considering embedded devices that are connected to a network offering the possibility to perform remote updates. A remote server can be used to update the binaries that are executed on the embedded device. We assume that the trustworthiness of the remote server is guaranteed (e.g. with methods as exemplified in \cite{Rauter2015}).


%

%

\subsection{Fault Model}
\begin{figure}[tb] 
  \centering
  \includegraphics[width=\columnwidth,keepaspectratio=true]{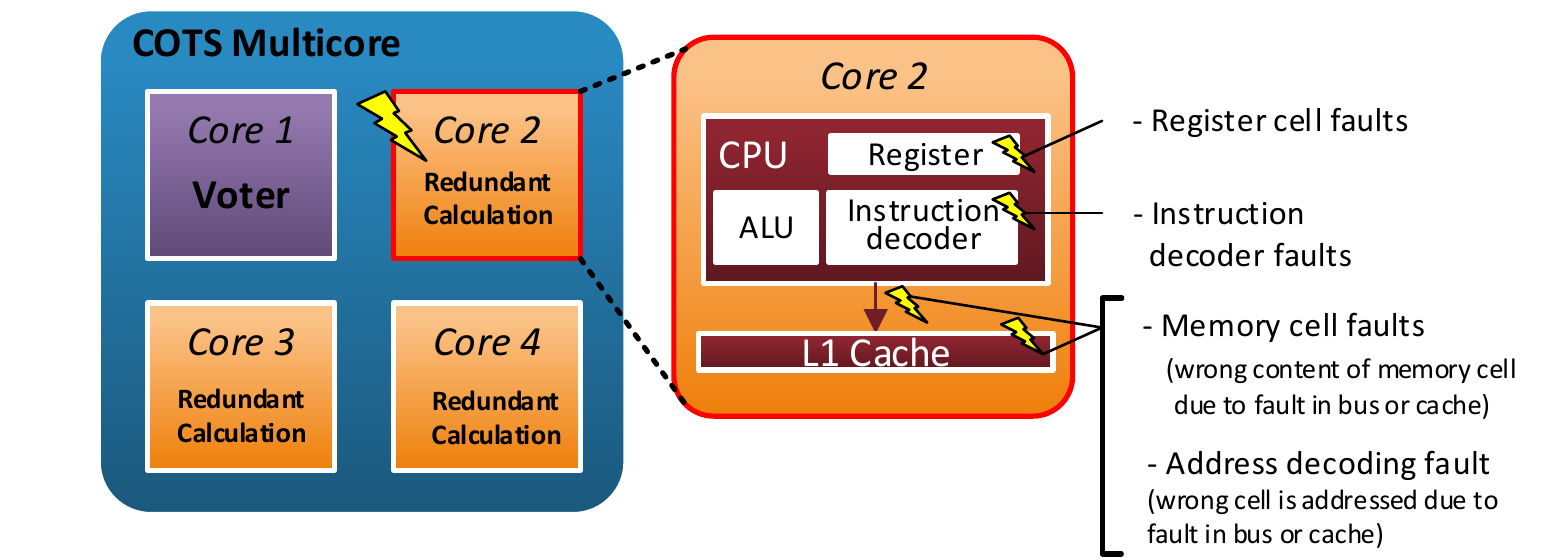}
  \caption{System overview and considered fault types. }
  \label{fig:faultmodel}  
\end{figure}

The scenario we address is the occurrence of a permanent fault in one of the cores executing a redundant calculation as shown in Fig. \ref{fig:faultmodel}. We consider fault sources of CPU-core elements as recommended by the IEC 61508 safety standard \cite{IEC61508}: registers, internal cache, instruction decoding, and address calculation/transfer.

\subsection{Adaptive Automated Software Diversity Concept Overview}
\begin{figure}[tb] 
  \centering
  \includegraphics[width=7cm,keepaspectratio=true]{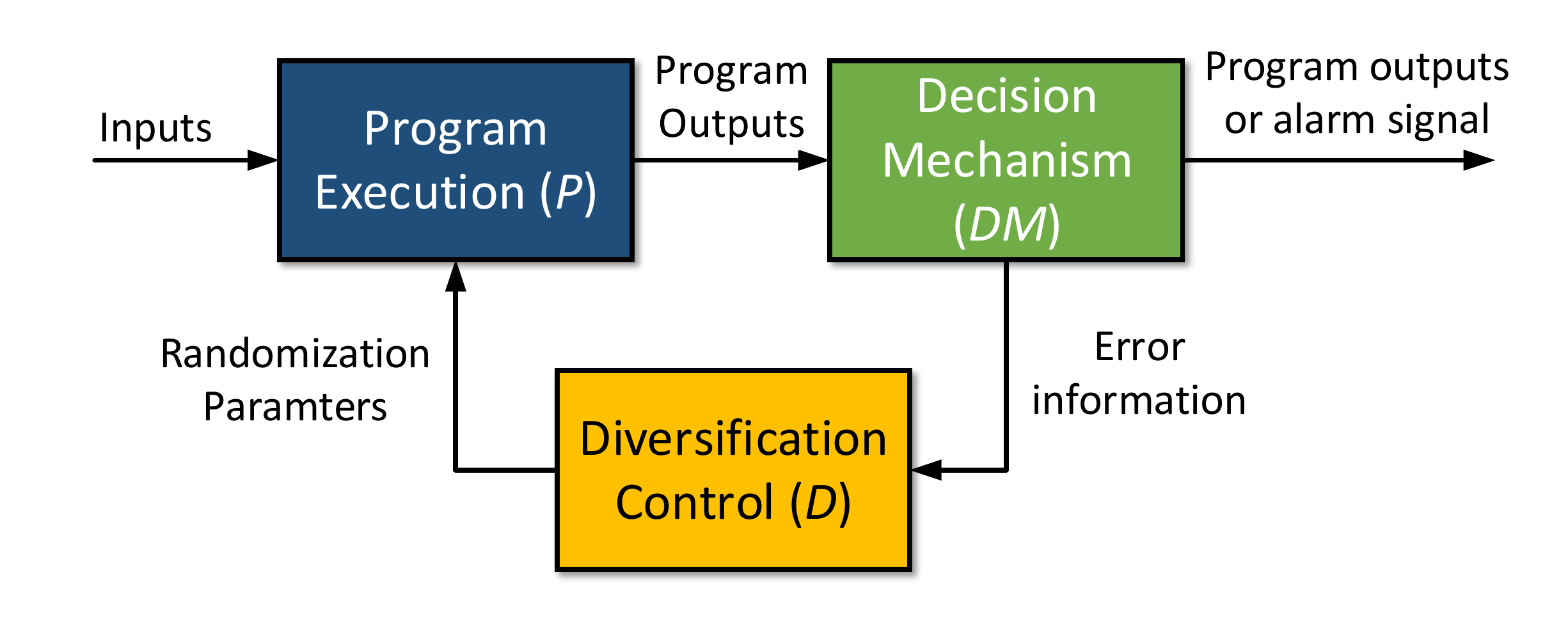}
  \caption{Basic structure of AASD. Based on information of a monitoring component (DM), a diversification controller decides whether and how to reconfigure the randomization mechanism of the main program \cite{Hoeller2015b}.
  }
  \label{fig:structure}
\end{figure}

To implement fault recovery, we propose to use the well-known concept of recovery blocks \cite{Pullum2001}. The main idea of recovery blocks is to execute alternate versions of a function, if the primary alternative of the function does not pass the acceptance test. Our approach extends this concept by generating the diverse replicas of the program automatically during runtime.
 This could be established by switching the statically generated program version or by changing the dynamic randomization. Fig. \ref{fig:structure} shows the basic principle of an adaptive automated software diversity (AASD) system \cite{Hoeller2015b}. Typically, a fault tolerant system contains the program, which performs the intended functionality of the system and a decision mechanism (DM) that monitors the program execution \cite{Pullum2001}. The DM detects anomalies, indicates alarms and decides which outputs to forward. In our application, the voter implements the DM. Additionally, we propose to add a component denoted as diversification control that creates a feedback-loop. This component manages the AASD by collecting and analyzing data on detected anomalies obtained from the DM. We propose to design the program in such a way that it can be randomized during execution according to parameters that can be adjusted during runtime (see Table \ref{tab:params}).
  
Then, the diversification control can decide to alter the execution by changing one or multiple parameters of the ASD configuration. 
 More specifically, we propose to learn from detected anomalies and to adapt the software of the replica that has been identified to be problematic by diversifying the execution with the above described ASD methods. 

\subsection{Fault Recovery Procedure}
\begin{figure}[tb] 
  \centering
  \includegraphics[width=\columnwidth,keepaspectratio=true]{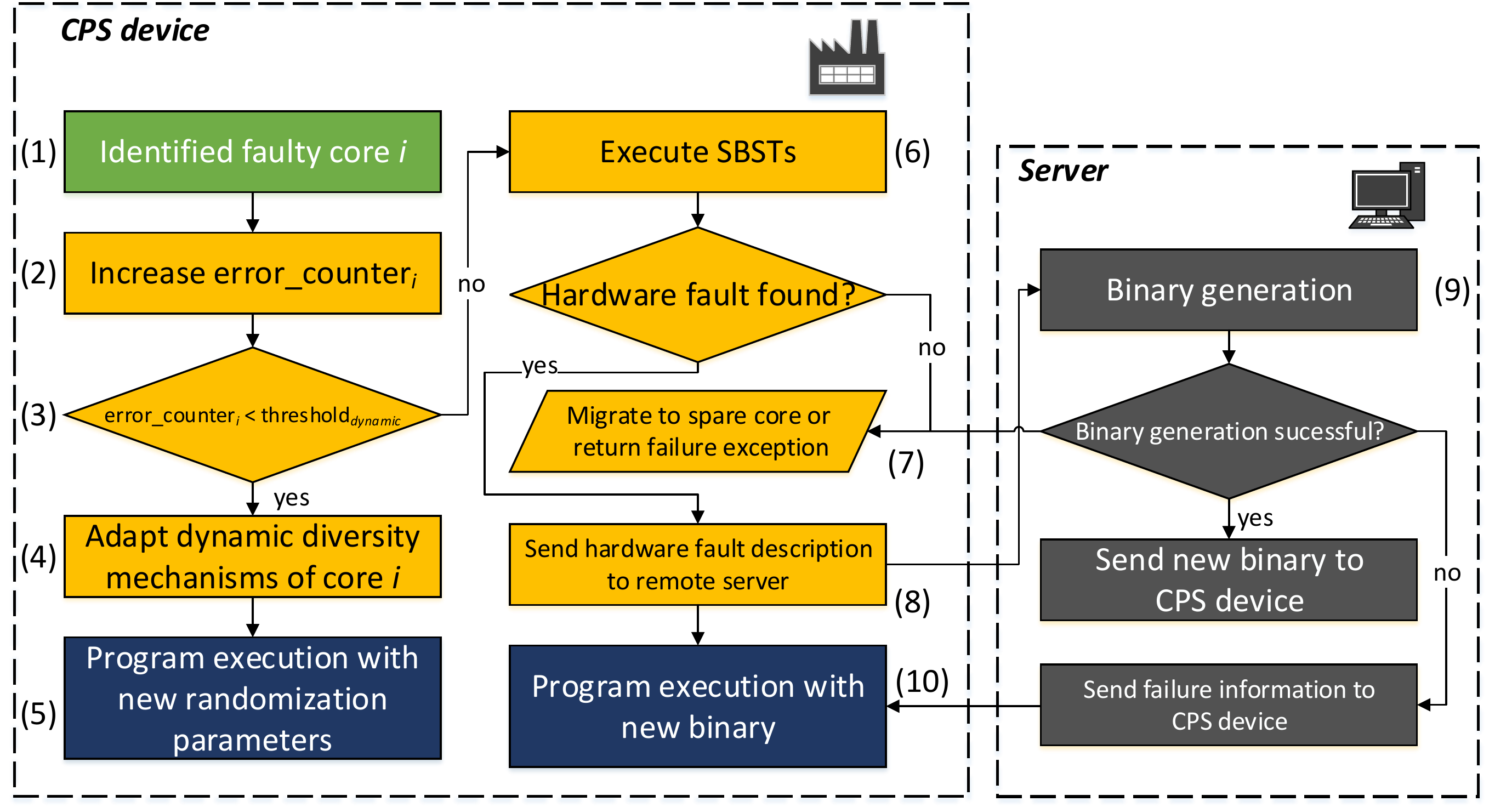}
  \caption{Basic procedure of fault recovery on a redundant multi-core system. The tasks colored in green are correspond to the decision mechanism, orange tasks are executed by the diversification control of Fig. \ref{fig:structure}.
  }
  \label{fig:flow}
\end{figure}

In Fig. \ref{fig:flow} an overview of the proposed fault recovery procedure is shown. If the voter detects a mismatch of the output values of the executions on the redundant core (1) an error counter corresponding to the faulty core is increased (2) . If subsequent executions of this core are performed correctly, it is assumed that the observed error has been caused by a transient fault and the error counter is set to zero. However, if more than a certain number of errors (\textit{threshold}$_{dynamic}$) have been observed, the core is regarded to be faulty and fault recovery approaches are applied (3). 

First, we try to adapt the execution with adaptive dynamic diversity methods (4)(5). Namely, the trend of the observed anomalies and the applied adjustable parameter settings of the dynamic diversity features are analyzed. Then, the dynamic diversity parameters are adjusted accordingly. In \cite{Hoeller2015serene}, we presented a basic example of how to implement such a mechanism. We assessed the method of introducing memory gaps with adjustable size. By changing this size during runtime the starting address of used variables can be changed. The size of the memory gaps represents the reconfiguration parameter that can be changed by the diversification control. Thus, if the diversification control decides to change a replica, the size of the memory gaps is adapted. Preliminary results indicate that this technique is effective in circumventing faulty memory regions. For example, for a simple bitcount application, the approach was able to bypass 94\% of introduced address decoder faults that would not have been masked without these mechanism. Table \ref{tab:dynamicfault} shows possible dynamic diversity techniques and the fault types, which we expect that could be tolerated.

\begin{table}
\caption{Expected fault types that can be addressed with example candidates of dynamic diversity techniques}
\begin{tabular}{L{3.2cm}C{0.8cm}C{0.8cm}C{0.8cm}C{0.9cm}}
\toprule
& \textit{Register cell fault} &
\textit{Instruction decoder fault} &
\textit{Memory cell fault}&
\textit{Address decoding fault}  \\ \midrule
Memory gaps  \cite{Bhatkar2005}                         &  &    & x & x   \\
Changing base addresses \cite{Bhatkar2005}                                      & x & x   \\
Data randomization \cite{Amman1988h}                                    & x &   & x &     \\
Memory layout randomization (e.g. \cite{Bhatkar2005,Chew2002})          &   &   & x & x   \\
Program encoding randomization (e.g. \cite{Barrantes2005})              &   & x &   &     \\ \bottomrule
\label{tab:dynamicfault}
\end{tabular}

\end{table}

If the adaptive dynamic diversity approach does not improve the fault tolerance, we try to recover from the fault with static diversity mechanisms. Therefore, first software-based self-tests are applied to identify the fault (6). We propose to also use software-based self-tests finding the considered fault types as described in the IEC 61508 safety standard. If the tests identify a fault, the fault definition (fault location, type of fault) is sent to a remote server (8). For example, these fault could be register bit with a value that cannot be changed (i.e. stuck-at-0 or stuck-at-1 faults). If no fault can be found, the fault recovery procedure is stopped and further fault handling can be applied (7). For example, the calculation can be mapped to a spare core, or an alarm is forwarded. Having the fault information and the source code, a powerful remote server tries to generate a variant of the software that bypasses the fault by using static diversity techniques (9). Details of this generation are described below. If the generation is successful, the embedded device receives the new binary and uses this binary for the execution on the faulty core (10). 


\subsection{Remote Binary Generation with Static Diversity Methods}
\begin{figure}[tb] 
  \centering
  \includegraphics[width=7cm,keepaspectratio=true]{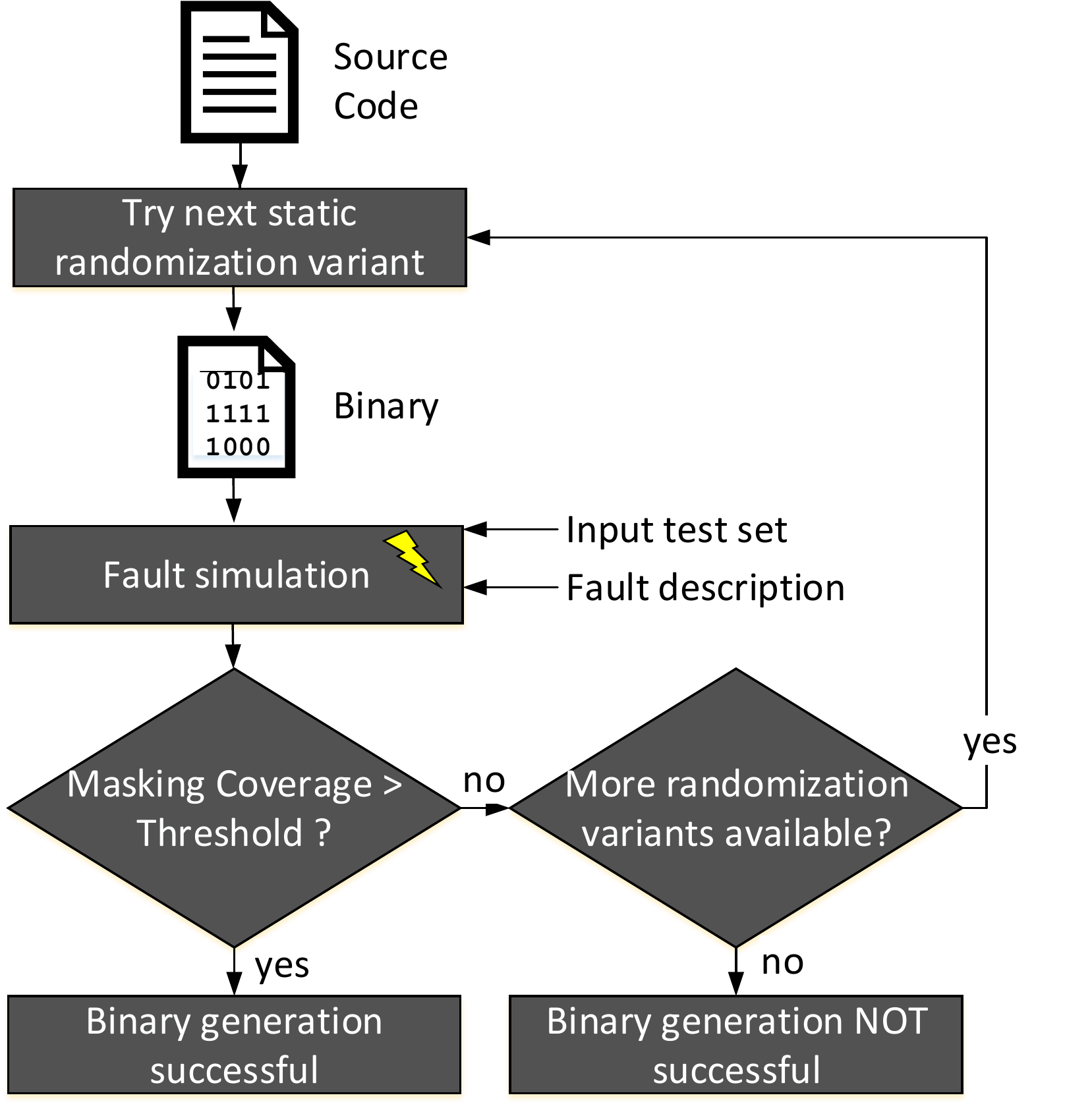}
  \caption{Basic procedure of binary generation on a remote server. Static diversity techniques are applied in order to get a binary that masks the given fault with a given probability (masking coverage). The figure shows details of the step (9) shown in Fig. \ref{fig:flow}. }
  \label{fig:server}
\end{figure}

Fig. \ref{fig:server} illustrates the binary generation procedure on the remote server. To generate diverse executions of the same source code static diversification methods are applied. Such methods could include source code mutations and using the compiler in such a way that the execution is diversified. As shown in \cite{Hoeller2015dasc}, the usage of diverse optimization flags during the compilation stage influences the amount of masked register and instruction decoder faults. We propose to exploit and leverage such effects in order to generate a binary that bypasses a given fault. For example, the GCC compiler allows to specify a register that should not be used (\texttt{-ffixed-reg}). If faults a certain register is faulty, this flag can be exploited to generate a binary that does not use this register. We propose to have multiple static diversity techniques in place. Then, different combinations of these diversity techniques can be applied to generate binaries. If a binary has been generated that masks the given fault with a certain probability (masking coverage) the binary generation procedure was successful and the binary is sent to the embedded device. 

We propose the following procedure to determine the probability that a given fault type is masked. To evaluate the consequences of a fault (i.e. fault masking, crash, or corrupted data) we propose to apply fault simulation with appropriate fault injection techniques. We use the QEMU-based fault injection framework as described in \cite{Hoeller2014mtv,Hoeller2015radiance}\cite{Hoeller2014mtv,Hoeller2015radiance}. This framework supports widely-deployed commercial off-the-shelf processor architectures and allows a software-based fault simulation without the need for software instrumentation. The framework features the injection of memory cell faults, address decoder faults, instruction decoder faults and register faults. Additionally, a input test set that represents typical input values is needed. Then, the processing of these input stimuli on a faulty hardware is simulated and the output is compared with a given golden reference output that would be generated on a correct hardware. We assume that the probability that the injected fault is masked can be approximated by the proportion of runs where the fault is masked and the total amount of tested input stimuli. 

\section{Research Challenges}
Although, there is a high fault recovery potential of the proposed technique, it poses many research challenges.
\paragraph{Determinism} Many dependable applications have fulfill real time requirements and thus require a deterministic execution. Here, the introduction of diversity would complicate the difficult enough task of determining the worst case execution time.

\paragraph{Implementation overhead} The impact on the fault detection latency and the resource requirements (e.g. performance, memory) should be kept low. 

\paragraph{Finding appropriate ASD techniques} We have presented potential ASD techniques above. However, studies are required to investigate in the feasibility of the approaches in a multi-core system. Furthermore, exhaustive evaluations on the effectiveness of the fault detection capabilities of the techniques for different kind of applications are required.

\paragraph{Adaptation tactics} Tactics of how to analyze the trends of detection malfunctions and ways of adapting the execution on redundant replicas have to be investigated.

\section{Conclusion}
There is a significant increase in complexity that threatens the system reliability. Complex embedded systems have to cope with an increasing number of hardware faults. Here, we proposed to use the opportunities inherent multi-core redundancy to achieve resilience. While typical research in this field only focuses on an homogeneous redundant execution, we propose to introduce diversity in the executed software versions. 

To keep the development overhead low and offer means to adapt the executions during runtime, we proposed to use automated software diversity approaches. Although, many promising approaches have been presented in literature, they have not been considered in the context of a redundant multi-core system so far. Thus, we tried to emphasize cross-fertilization of security, reliability, and multi-core research. We highlighted the idea of using automated software diversity techniques for fault recovery. 

In the future, we plan to evaluate adaptive automated software diversity techniques for specific complex applications in a multi-core system by implementing a prototype. Furthermore, we hope to encourage further researchers to explore techniques based on this promising yet challenging idea.

\bibliographystyle{IEEEtran}
\bibliography{library}

\balance

\end{document}